\documentclass[12pt]{article}
\usepackage{a4wide}
\usepackage{graphicx}

\usepackage{amsbsy}

\usepackage{amsmath, amsthm, amssymb}

\usepackage{dsfont}

\usepackage{mathrsfs}

\begin{document}
{\renewcommand{\thefootnote}{\fnsymbol{footnote}}
\hfill  IGC--09/6--2\\
\medskip
\begin{center}
{\LARGE  Effective Constraints for Relativistic Quantum Systems}\\
\vspace{1.5em}
Martin Bojowald\footnote{e-mail address: {\tt bojowald@gravity.psu.edu}}
and Artur Tsobanjan\footnote{e-mail address: {\tt axt236@psu.edu}}
\\
\vspace{0.5em}
Institute for Gravitation and the Cosmos,
The Pennsylvania State
University,\\
104 Davey Lab, University Park, PA 16802, USA\\
\vspace{1.5em}

\end{center}
}

\begin{abstract}
 Determining the physical Hilbert space is often considered the most
 difficult but crucial part of completing the quantization of a
 constrained system. In such a situation it can be more
 economical to use effective constraint methods, which are extended
 here to relativistic systems as they arise for instance in quantum
 cosmology. By side-stepping explicit constructions of states, such
 tools allow one to arrive much more feasibly at results for physical
 observables at least in semiclassical regimes. Several questions
 discussed recently regarding effective equations and state properties
 in quantum cosmology, including the spreading of states and quantum
 back-reaction, are addressed by the examples studied here.
\end{abstract}

\section{Introduction}

One of the key issues in quantizations of fundamental theories, which
due to their covariance properties are systems with gauge freedom
generated by constraints, is the determination of physical
observables. They must satisfy the constraint equations and be
invariant under gauge transformations. For canonical quantum theories,
solving constraints is traditionally done at the state level: one
constructs a physical Hilbert space of states annihilated by the
constraint operator(s) and equipped with an invariant inner
product. Explicit constructions can be done in some special cases by
different methods.

Since explicit derivations are possible only in specific cases, it
is not always clear whether the results are generic or mere
artefacts of the simple models used. It is therefore important to
have approximate methods for a wider range of cases, or at least to
be able to perturb around known solvable ones while still ensuring
that the constraints are solved and the observables are gauge
invariant. It turns out that such perturbation schemes are most
feasible if one deals with the observables directly, such as
expectation values, side-stepping the computation and physical
normalization of states. This procedure gives rise to canonical
effective equations and constraints \cite{EffAc,EffCons}.

A procedure for effective constraints has been formulated in
\cite{EffCons} and applied to parameterized non-relativistic systems
with a constraint $p_t+H=0$ where $p_t$ is the momentum of time and
$H$ the Hamiltonian. (The concepts and results are reviewed briefly
below.) It was shown that the physical observables in suitable
regimes, including semiclassical ones, can be determined without
making use of a physical inner product but instead through
implementing reality conditions for quantum variables such as
fluctuations, correlations and higher moments. For applications of
these methods to quantum gravity and cosmology one has to extend
them to relativistic systems, offering one additional subtlety:
constraints would now be of the form $p_t^2-H^2=0$, requiring one to
take a square root and to make sign choices. Mathematically, for
instance, the question is how to precisely define $\sqrt{H^2}=|H|$
at the operator level. This may not be obvious if the operator
$\mathbf{H}$ has a complicated spectrum or is not positive definite.
Physically, one must decide how to treat and separate positive and
negative frequency solutions corresponding to the two solutions
$p_t=\pm|H|$. (See e.g.\ \cite{ParticleWaveMechanics,GenRepIn} for
discussions of relativistic systems.)

For a time-independent Hamiltonian, it turns out that one can, at
least for semiclassical purposes, simply use
$p_t=\pm\langle\mathbf{H}\rangle$ as the effective Hamiltonian
\cite{BouncePert,BounceCohStates} without absolute values, even if
$\mathbf{H}$ is not positive definite. One only has to ensure that
the initial values used in the effective equations of motion
correspond to an initial state supported on a part of the spectrum
of $\mathbf{H}$ with a definite sign. On such a state and with a
self-adjoint $\mathbf{H}$, $\langle|\mathbf{H}|\rangle=
\langle\mathbf{H}\rangle$ if the state is supported only on the
positive part of the spectrum of $\mathbf{H}$, and
$\langle|\mathbf{H}|\rangle= -\langle\mathbf{H}\rangle$ if the state
is supported only on the negative part.  Since the Hamiltonian is
preserved, these statements will hold true throughout the whole
evolution and there is no need for an absolute value in the
effective Hamiltonian. This fact has been made use of in several
recent derivations of effective equations in quantum cosmology,
where the relevant versions of $\mathbf{H}$ are not positive
definite \cite{Harmonic,Recollapse}.

In those models, deparameterization was performed using $t=\phi$ as
an internal time from a free, massless scalar $\phi$. The same types
of models also allow the construction of a physical Hilbert space at
the state level \cite{Blyth,APS}, with the results in agreement with those
obtained from effective equations.  Most interesting from the
cosmological perspective is, however, a system where the scalar
$\phi$ has a non-trivial potential or at least a mass term. This has
two immediate implications: in general, one can no longer
deparameterize globally since the solutions for the scalar would not
be monotonic in the time coordinates, and the Hamiltonian would no
longer be (internal) time independent. Positivity can no longer be
ensured just by an initial condition, and using
$p_{\phi}=\pm\langle\mathbf{H}(\phi)\rangle$ as an effective
Hamiltonian without an absolute value may then seem questionable.
Explicitly using an absolute value, on the other hand, would make a
derivation of the effective equations more complicated. At this
stage, a direct treatment of effective constraints for relativistic
systems without deparameterization becomes relevant. This is what we
present in the current paper.

We will consider in detail models of relativistic particles and
properties of observables in their quantum theories. In the massive
case, for instance, we are dealing with the quantization and
implementation of a constraint $C=p_t^2-p^2-m^2$. For physical
states, the expectation value
\[
 \langle\mathbf{C}\rangle= p_t^2+(\Delta
p_t)^2-p^2-(\Delta p)^2-m^2
\]
must vanish and thus represents a quantum constraint. (We will
notationally identify classical degrees of freedom with the
expectation values to simplify the notation and to show the relation
between classical terms and quantum corrections. Thus,
$E=\langle\mathbf{E}\rangle$ and $(\Delta E)^2=
\langle(\mathbf{E}-\langle\mathbf{E}\rangle)^2\rangle=
\langle\mathbf{E}^2\rangle- E^2$.) As we will see in more detail
below, there are additional independent constraints since
expressions such as $\langle\mathbf{q}\mathbf{C}\rangle$ must also
vanish for physical states. Allowing all possible factors to the
left of $\mathbf{C}$, this presents a constrained system of
infinitely many constraints for infinitely many quantum variables
given by the moments of a state. The combined system of all
constraints must be solved to find observable results, which is
feasible in semiclassical regimes where only a finite set of moments
suffices to characterize a state approximately. The same kind of
approximation also allows one to include potentials within the
constraint, which may be explicitly time-dependent. We will exploit
this to justify the procedures used in quantum cosmology for
deparameterized systems with time-dependent potentials, as developed
in \cite{BouncePot,QuantumBounce,BounceSqueezed}.

Another question of interest is that of the spreading of states and
quantum back-reaction of fluctuations and higher moments on the
expectation values. If we compare the effective constraint
$\langle\mathbf{C}\rangle$ written above, which contains only the
second order moments in addition to the expectation values, with the
effective Hamiltonian of the corresponding deparameterized system,
\begin{eqnarray}
 \langle\mathbf{H}\rangle&=& \langle\sqrt{\mathbf{p}^2+m^2}\rangle=
 \langle\sqrt{(p+(\mathbf{p}-p))^2+m^2}
 =\sqrt{p^2+m^2}+\sum_{n=2}^{\infty} \frac{1}{n!} \frac{\partial^n
 \sqrt{p^2+m^2}}{\partial p^n}\langle(\mathbf{p}-p)^n\rangle\nonumber\\
&=& \sqrt{p^2+m^2}+\frac{m^2(\Delta p)^2}{(p^2+m^2)^{3/2}} -3m^2
\frac{m^2-4p^2}{(p^2+m^2)^{7/2}}\langle(\mathbf{p}-p)^3\rangle +
\cdots \label{EExp}
\end{eqnarray}
with a whole formal Taylor series that includes higher moments,
different coupling terms between the expectation values and the
moments seem to be implied. Then, back-reaction might seem different
in these two treatments, apparently making them incompatible. By our
specific constructions in this paper we will reconcile these apparent
disagreements, and provide an illustration by numerical solutions in a
specific example (App.~\ref{AppII}). This is also important for
quantum cosmology, where quantum back-reaction is crucial to the
understanding of how a quantum state evolves toward and possibly
through the big bang and how much of the pre-big bang state can be
reconstructed \cite{BeforeBB,RecallComment}.

\section{Effective constraints}

In a canonical effective description, the dynamics of a quantum
system with $n$ degrees of freedom is formulated in terms of the
expectation values, i.e.\ the evaluations of a state functional in
the elements of an algebra $\mathscr{A}$ generated by $2n$ basic
operators $\mathbf{q}_i$ and $\mathbf{p}_i$, $i=1,\ldots,n$.  This
whole set of infinitely many variables can be conveniently split
into the $2n$ expectation values of the basic operators, such as
$q_i=\langle\mathbf{q}_i\rangle$ and
$p_i=\langle\mathbf{p}_i\rangle$, $i=1,\ldots,n$, together with the
infinitely many moments of the form
\[
G^{\{a_j\},\{b_j\}}=\left\langle\prod_{i=1}^n(\mathbf{q}_i-\langle \mathbf{q}_i
\rangle)^{a_i} (\mathbf{p}_i - \langle \mathbf{p}_i \rangle)^{b_i}
\right\rangle_{\rm Weyl}
\]
where the subscript ``Weyl'' denotes the totally symmetric ordering
of all factors involved. Between all these variables a Poisson
bracket is defined following from the algebra of commutation relations:
\begin{equation}
\{\langle\mathbf{A}\rangle,\langle\mathbf{B}\rangle\}=
\frac{\langle[\mathbf{A},\mathbf{B}]\rangle}{i\hbar}
\end{equation}
extended using linearity and the Leibniz rule.

At second order, $\sum_{i=1}^n (a_i+b_i)=2$, this set of moments
includes all fluctuations and covariances. In semiclassical regimes,
moments fall off at least as
$\hbar^{\frac{1}{2}\sum_{i=1}^n(a_i+b_i)}$ such that only low orders
need to be considered for the first approximation to quantum
effects. Below we will employ the notation $(\Delta a)^2 = \langle
(\mathbf{a}-a)^2 \rangle$ for fluctuations and $\Delta(ab)=
\langle(\mathbf{a}-a)(\mathbf{b}-b)\rangle_{\rm
Weyl}=\frac{1}{2}\langle(\mathbf{a}-a)(\mathbf{b}-b)+
(\mathbf{b}-b)(\mathbf{a}-a)\rangle $ for covariances, which may
make second order equations easier to interpret.

Any operator $\mathbf{C}=C(\mathbf{q}_i,\mathbf{p}_j)$ gives rise to a
function on the space of states, which can be expressed in terms of the
moments by Taylor expanding
\begin{equation} \label{QuantCons}
 \langle C(\mathbf{q}_i,\mathbf{p}_j)\rangle=
\langle C(q_i+(\mathbf{q}_i-q_i),p_j+(\mathbf{p}_j-p_j))\rangle
\end{equation}
in $\mathbf{q}_i-q_i$ and $\mathbf{p}_j-p_j$ as in (\ref{EExp}). If
$C(\mathbf{q}_i,\mathbf{p}_j)$ is a constraint operator,
(\ref{QuantCons}) must vanish on physical states and thus is a
constraint on the quantum phase space.

A single constraint on the phase space removes one pair of
variables, but not the whole tower of moments associated with it in
the quantum phase space. For a complete reduction one has to make
use of additional constraints, provided by the set of phase space
functions $\langle
f(\mathbf{q}_i,\mathbf{p}_i)C(\mathbf{q}_j,\mathbf{p}_k)\rangle$
which must also vanish on physical states. These functions are in
general independent from the quantum constraint $\langle
C(\mathbf{q}_i,\mathbf{p}_j)\rangle$ as functions of expectation
values and moments. As shown in \cite{EffCons}, this set of
constraints remains first class. (Note that we do not order the
products of the operators in the constraints symmetrically to ensure
that the constraint operator acts directly on the state. As a
result, in some cases one has to deal with complex-valued
constraints requiring reality conditions for physical observables.
This has been discussed in \cite{EffCons} and will also be seen in
more detail in the examples below.)

Given such a system of constraints on the quantum phase space, one
can proceed in the classical way and find the reduced quantum phase
space of observables or solve the constraints and fix the gauge. At
this stage, it is convenient (but not required) to decide which
internal time variables $(t,p_t)$ among the $(q_i,p_j)$ should be
used. Since a quantization of the corresponding deparameterized
system, if it exists, would not give rise to any moment involving an
operator of $t$ or $p_t$, solving the constraints must remove all
moments including at least one factor of $\mathbf{t}$ or
$\mathbf{p}_t$ from the original quantum phase space. That this
indeed happens was verified to second order of the parameterized
non-relativistic particle in \cite{EffCons}. In that case, there was
a single linear term $p_t$ in the constraint, such that all
$p_t$-moments $(\Delta p_t)^2$, $\Delta(p_tt)$, $\Delta (p_tq)$ and
$\Delta(p_tp)$ can be removed by solving the system of constraints,
to second order:
\begin{eqnarray*}
 \langle(\mathbf{q}-q)\mathbf{C}\rangle&=& \Delta(qp_t)+
 \frac{p}{M}\frac{i\hbar}{2}+
\frac{p}{M} \Delta(qp)= 0\\
 \langle(\mathbf{t}-t)\mathbf{C}\rangle&=&\frac{p}{M} \Delta(pt)+
\Delta(tp_t)+\frac{i\hbar}{2}= 0\\
 \langle(\mathbf{p}_t-p_t)\mathbf{C}\rangle&=&(\Delta p_t)^2+\frac{p}{M} \Delta(pp_t)
= 0\\
 \langle(\mathbf{p}-p)\mathbf{C}\rangle&=&\Delta(pp_t)+\frac{p}{M} (\Delta p)^2
= 0\,.
\end{eqnarray*}
This leaves the moments $(\Delta t)^2$, $\Delta (tq)$ and $(\Delta
tp)$, which are removed by factoring out the gauge flow, or simply by
setting them to zero as a well-defined set of gauge-fixing
conditions. (Note that a smaller number of gauge fixing conditions
than constraints is required because the second order moments satisfy
a Poisson algebra which is degenerate from the symplectic point of
view; see \cite{brackets} for some notions of constrained systems in the
non-symplectic case. Additionally, setting the fluctuation $(\Delta
t)^2$ to zero is consistent with the generalized uncertainty relation
\[
 (\Delta t)^2(\Delta p_t)^2- \Delta(tp_t)^2\geq \frac{\hbar^2}{4}
\]
since $t-p_t$-correlations $\Delta(tp_t)=-i\hbar/2$ are required by
the constraints, especially
$\langle(\mathbf{t}-t)\mathbf{C}\rangle=0$ with the gauge fixing
condition $\Delta(pt)=0$, to be imaginary and of just the right size
to saturate the uncertainty relation.) After solving the constraints
and fixing the gauge, observable moments are recovered on which
physical reality conditions can be imposed.

\section{Free relativistic particle}
\label{sec:free_particle}

Classically a free relativistic particle is described by a single
constraint
\[
C = p_t^2 - p^2 - m^2
\]
on the phase space coordinatized by two canonical pairs $t, p_t$\
and $q, p$.\footnote{We assume the units have been chosen so that
both length and momentum have the units of the square root of action
(e.g. geometrized units).} For the quantum version we consider the
unital associative algebra generated by four basic elements
$\mathbf{t}, \mathbf{p}_t, \mathbf{q}, \mathbf{p}$\ subject to the
canonical commutator relations. That is, $\mathscr{A}$\ consists of
(countable) sums of polynomials of the form $\mathbf{t}^i
\mathbf{p}_t^j \mathbf{q}^k \mathbf{p}^l$; terms with a different
ordering may be expressed using the commutation relations
\[
[\mathbf{t}, \mathbf{p}_t] = i\hbar \mathds{1} \quad, \quad
[\mathbf{q}, \mathbf{p}] = i \hbar \mathds{1}\,.
\]
There is no product-ordering ambiguity in the case of the above
constraint and it is naturally identified with an element
$\mathbf{C} = \mathbf{p}_t^2 - \mathbf{p}^2 - m^2 \mathds{1}$\ of
$\mathscr{A}$. As the equivalent of Dirac's condition $\mathbf{C}
\psi = 0$\ we demand that the constraint has a vanishing right
action on the states, which in our case are complex linear functions
$\alpha : \mathscr{A} \rightarrow \mathds{C}$, this implies that
$\alpha(\mathbf{aC}) = 0, \ \forall \mathbf{a} \in \mathscr{A}$;
henceforth we drop explicit reference to $\alpha$\ and write this
condition as $\langle \mathbf{aC} \rangle = 0$ for the expectation
value in a physical state $\alpha$. In order to impose all these
conditions systematically we take the previously mentioned basis in
$\mathscr{A}$\ and impose the constraint via an infinite (but
countable) set of conditions
\begin{equation}
\langle \mathbf{t}^i \mathbf{p}_t^j \mathbf{q}^k \mathbf{p}^l
\mathbf{C} \rangle = 0 \label{eq:constraint_functions}\,.
\end{equation}

We reduce the above infinite system of equations using the same
method that was previously employed for a Newtonian
particle~\cite{EffCons}---a semiclassical expansion based on the
hierarchy
\[
\left\langle(\mathbf{t}-\langle \mathbf{t} \rangle)^{i}
(\mathbf{p}_t-\langle \mathbf{p}_t \rangle)^{j} (\mathbf{q}-\langle
\mathbf{q} \rangle)^{k} (\mathbf{p} - \langle \mathbf{p}
\rangle)^{l} \right\rangle_{\rm Weyl} \propto
\hbar^{\frac{1}{2}(i+j+k+l)}
\]
of moments.

\subsection{Constraints at second order}
\label{sec:massive_particle_constraints}

In what follows, we assume a semiclassical state and drop the terms of
order $\hbar^{\frac{3}{2}}$\ and above, keeping the terms of order
$\hbar$\ and below. This will suffice to demonstrate the feasibility
of our methods for relativistic systems. To this order our system is
described by fourteen independent functions: four expectation values
of the form $a= \langle \mathbf{a} \rangle$; four spreads of the form
$(\Delta a)^2 = \langle (\mathbf{a}-a)^2 \rangle$\ and six covariances
of the form $\Delta(ab)=
\langle(\mathbf{a}-a)(\mathbf{b}-b)\rangle_{\rm Weyl}$. (Poisson
brackets between all these variables are listed in App.~\ref{AppI}.) The
infinite system of constraint functions is reduced to just five
non-trivial conditions
\begin{eqnarray}
C &=& \langle \mathbf{C} \rangle = p_t^2 - p^2 - m^2 + (\Delta
p_t)^2 - (\Delta p)^2 = 0 \nonumber \\ C_{t} &=& \langle
(\mathbf{t}-\langle \mathbf{t} \rangle) \mathbf{C} \rangle = 2p_t
\Delta(t p_t) + i\hbar p_t - 2p \Delta(t p) = 0 \nonumber \\
C_{p_t} &=& \langle (\mathbf{p}_t-\langle \mathbf{p}_t \rangle)
\mathbf{C}
\rangle = 2p_t (\Delta p_t)^2 - 2p \Delta(p_tp) = 0 \nonumber \\
C_q &=& \langle (\mathbf{q}-\langle \mathbf{q} \rangle) \mathbf{C}
\rangle = 2p_t \Delta(p_tq) - 2p \Delta(qp) - i\hbar p = 0 \nonumber
\\ C_p &=& \langle (\mathbf{p}-\langle \mathbf{p} \rangle) \mathbf{C}
\rangle = 2p_t \Delta(p_tp) - 2p(\Delta p)^2 = 0\,.
\label{eq:2_ord_constr_m}
\end{eqnarray}
Note that the semiclassical hierarchy of variables is critical to
the above reduction in the number of constraint conditions. In
particular, $C=0$\ implies that $p_t^2 - p^2-m^2 = (\Delta p)^2 -
(\Delta p_t)^2$, which in turn implies that the combination of the
expectation values $C_{\rm Class}:=p_t^2 - p^2-m^2$\ is of order
$\hbar$\ on the constraint surface. In other words, the classical
constraint is satisfied to order $\hbar$. The terms of the form
$(\Delta a)^2C_{\rm Class}$\ and $\Delta(ab) C_{\rm Class}$\ are
then of order $\hbar^2$\ and should be dropped in our present
treatment. The complete infinite system of constraint functions is a
closed Poisson algebra---or, in the language of classical constraint
analysis, a \emph{first-class} system~\cite{DirQuant}. In general,
due to the nature of the above truncation one would expect the
reduced system of constraints to remain closed only to order
$\hbar$. In our case the Poisson algebra of the truncated set of
constraint functions---displayed in Table~\ref{tab:pb_m}---is
\emph{exactly} closed with respect to the bracket.

\begin{table}[ht]
\centering \caption{Poisson algebra of constraints for a free
particle. First terms in the bracket are labeled by rows, second
terms are labeled by columns.}
\begin{tabular}{|c|| c|c|c|c|c|}
\hline & ${\scriptstyle C}$ & ${\scriptstyle C_{t}}$ &
${\scriptstyle C_{p_t}}$ & ${\scriptstyle C_q}$ & ${\scriptstyle
C_p}$  \\ \hline\hline ${\scriptstyle C}$ & ${\scriptstyle 0}$ &
${\scriptstyle -2C_{p_t}}$ & ${\scriptstyle 0}$ & ${\scriptstyle
-2C_p}$ & ${\scriptstyle 0}$
\\ \hline ${\scriptstyle C_{t}}$ & ${\scriptstyle 2C_{p_t}}$ &
${\scriptstyle 0}$ & ${\scriptstyle 4p_tC_{p_t}-2pC_p}$ &
${\scriptstyle 2pC_{t} + 2p_tC_q}$ & ${\scriptstyle 2p_tC_p}$
\\ \hline ${\scriptstyle C_{p_t}}$ & ${\scriptstyle 0}$ &
${\scriptstyle 2pC_p-4p_tC_{p_t}}$ & ${\scriptstyle 0}$ &
${\scriptstyle 2pC_{p_t}}$ & ${\scriptstyle 0}$ \\ \hline
${\scriptstyle C_q}$ & ${\scriptstyle 2pC_p}$ & ${\scriptstyle
-2pC_{t}-2p_tC_q}$ & ${\scriptstyle -2pC_{p_t}}$ & ${\scriptstyle
0}$ & ${\scriptstyle 2p_tC_{p_t}-4pC_p}$ \\ \hline ${\scriptstyle
C_p}$ & ${\scriptstyle 0}$ & ${\scriptstyle -2p_tC_p}$ &
${\scriptstyle 0}$ & ${\scriptstyle 4pC_p - 2p_tC_{p_t}}$ &
${\scriptstyle 0}$ \\ \hline
\end{tabular}
\label{tab:pb_m}
\end{table}

To solve the constraint system we eliminate five variables using the
five conditions from~(\ref{eq:2_ord_constr_m}). Specifically, we
eliminate the five quantum variables associated with $\mathbf{p}_t$,
having in mind that $t$ will be chosen as time in a deparameterized
treatment.  We start by noting that $p_tC_{p_t}=0$\ gives
\begin{equation} \label{ECE}
0 = p_t^2 (\Delta p_t)^2 - pp_t \Delta(p_tp)\,.
\end{equation}
However $C_p=0$\ implies
\[
p_t\Delta(p_tp) = p (\Delta p)^2
\]
and substituted in (\ref{ECE}) gives
\[
0=p_t^2 (\Delta p_t)^2 - p^2 (\Delta p)^2\,.
\]
Finally, eliminating $p_t^2$\ through $C=0$\ we obtain a quadratic
equation in $(\Delta p_t)^2$
\[
\left( (\Delta p_t)^2 \right)^2 - (\Delta p_t)^2(p^2 + m^2 + (\Delta
p)^2) + p^2 (\Delta p)^2 = 0
\]
with two solutions
\[
(\Delta p_t)^2 = \frac{1}{2} \left( p^2 + m^2 + (\Delta p)^2 \pm
\sqrt{ (p^2 + m^2 + (\Delta p)^2)^2 - 4p^2(\Delta p)^2 } \right)\,.
\]
In order to see whether either solution is compatible with the
hierarchy assumed by the semiclassical approximation, we expand the
solution to order~$\hbar$. One finds
\begin{eqnarray*}
(\Delta p_t)^2 &=& \frac{1}{2} (p^2 + m^2) \left( 1 + \frac{(\Delta
p)^2}{p^2 + m^2} \pm \sqrt{1 + \frac{(\Delta p)^2(2m^2 - 2p^2 +
(\Delta p)^2)}{(p^2 + m^2)^2}} \right) \\ &=& \frac{1}{2} (p^2 +
m^2) \left( 1 \pm 1 + \frac{(\Delta p)^2}{(p^2 + m^2)^2} (p^2 + m^2
\pm (m^2 -p^2)) + O\left((\Delta p)^4\right) \right)\,.
\end{eqnarray*}
Looking at the solution with the ``$+$'' sign we see the following
leading order behavior
\[
(\Delta p_t)^2 = p^2 + m^2 + O(\hbar)
\]
which is inconsistent with the assumption that $(\Delta p_t)^2$\ is
of order $\hbar$. The solution with the ``$-$'' sign leads to
\[
(\Delta p_t)^2 =\frac{p^2(\Delta p)^2}{p^2 + m^2} + O\left((\Delta
p)^4\right)
\]
which is of order $\hbar$\ and therefore consistent with the
semiclassical approximation. Substituting the latter solution back
into the constraint conditions~(\ref{eq:2_ord_constr_m}) we obtain
two sets of solutions
\begin{eqnarray}
p_t &=& \pm E \nonumber \\
\Delta(tp_t) &=& \pm \frac{p}{E} \Delta(tp) - \frac{i\hbar}{2}
\nonumber \\ (\Delta p_t)^2 &=& p^2 + m^2 + (\Delta p)^2 - E^2
\nonumber \\ \Delta(p_tq) &=& \pm \frac{p}{E} \left( \Delta(qp) +
\frac{i\hbar}{2} \right) \nonumber \\ \Delta(p_tp) &=& \pm
\frac{p}{E} (\Delta p)^2 \label{eq:2nd_ord_sol_m}
\end{eqnarray}
where
\[
E = \frac{1}{\sqrt{2}} \left( p^2 + m^2 + (\Delta p)^2
+  \sqrt{ (p^2 + m^2 + (\Delta p)^2)^2 - 4p^2(\Delta p)^2 }
\right)^{\frac{1}{2}} \,.
\]
By rearranging the terms in the above expression one can easily
verify that $E \geq 0$\ when reality and positivity of the physical
variables are imposed (see
Section~\ref{sec:massive_particle_gauge}). One recovers the usual
relativistic dispersion relation where energy equals $\sqrt{p^2 +
m^2}$\ if one assumes a ``momentum eigenstate'', that is if one sets
the spread in momentum $(\Delta p)^2 = 0$.

\subsection{Gauge freedom}\label{sec:massive_particle_gauge}

The truncated system of constraints~(\ref{eq:2_ord_constr_m}) is
equivalent (when consistency with the semiclassical approximation is
evoked) to the restriction to two disjoint surfaces, each one
corresponding to a choice of sign in~(\ref{eq:2nd_ord_sol_m}). Each
surface is described by an equivalent set of ``linearized''
constraints
\begin{eqnarray}
C_{1\pm} &=&  p_t \pm E \nonumber \\ C_{2\pm} &=& (\Delta
p_t)^2 - p^2 - m^2 - (\Delta p)^2 + E^2 \nonumber \\
C_{3\pm}
&=& \Delta(p_tp) \pm \frac{p}{E}(\Delta p)^2 \nonumber \\
C_{4\pm} &=& \Delta(p_tq) \pm \frac{p}{E} \left( \Delta(qp) +
\frac{i\hbar}{2} \right) \nonumber \\ C_{5\pm} &=& \Delta(tp_t) \pm
\frac{p}{E} \Delta(t p) + \frac{i\hbar}{2}\,.
\end{eqnarray}
The above constraint conditions can be expressed as sums of the
conditions in~(\ref{eq:2_ord_constr_m}) and therefore form a first
class system. Additionally, for the calculations to follow it is
useful to note that $p_t$, $p$, $(\Delta p_t)^2$, $\Delta(p_tp)$ and
$(\Delta p)^2$\ are first class functions with respect to either set
of constraints. This can ultimately be traced back to the fact that
$[\mathbf{p}_t, \mathbf{C}] = 0 = [\mathbf{p}, \mathbf{C}]$. It
follows that $E$, which is a function of $p$\ and $(\Delta p)^2$\
only, is also first class.

On the constraint surfaces the ``linearized'' constraints can be
used to eliminate the five variables $p_t$, $(\Delta p_t)^2$,
$\Delta(tp_t)$, $\Delta(p_tp)$, $\Delta(p_tq)$. At this stage there
remain four degrees of freedom associated with the algebra elements
generated by $\mathbf{t}$. These will be treated as gauge parameters
associated with the time evolution of the system. From this point of
view, one has a four-parameter space to choose from when it comes to
the evolution of the ``physical variables'' (i.e. those associated
with the algebra generated by $\mathbf{q}$\ and $\mathbf{p}$\
alone). Viewing our system expanded to second order in quantum
variables as a classical constrained system, the evolution on the
``physical variables'' --- $q$, $p$, $(\Delta q)^2$, $\Delta(qp)$,
$(\Delta p)^2$
--- may be generated by any constraint function of the form
\begin{equation}
C_{\rm Ham} = \sum_{i} \mu_i C_{i\pm}  \label{eq:C_ham}
\end{equation}
where the multipliers $\mu_i$\ are arbitrary functions of the physical
variables. The presence of several constraints, all associated with
the classical Hamiltonian, means that a priori there is no unique time
parameter. Depending on the choice of gauge, any combination of $t$
with moments involving $\mathbf{t}$ can play the role of time.

At this stage we would like to restrict the gauge freedom
down to a single parameter and to interpret the first class flow
in the direction of $t$\ as the dynamical evolution of the
system. This may be accomplished by introducing three gauge choices
\begin{eqnarray*}
\phi_1 &=& (\Delta t)^2 - f_1(q, p, (\Delta q)^2, \Delta(qp),
(\Delta p)^2)=0 \\ \phi_2 &=& \Delta(tq) - f_2(q, p, (\Delta q)^2,
\Delta(qp), (\Delta p)^2) = 0 \\ \phi_3 &=& \Delta(tp) - f_3(q, p,
(\Delta q)^2, \Delta(qp), (\Delta p)^2) = 0
\end{eqnarray*}
with functions $f_1$, $f_2$ and $f_3$ to be determined.  We define
$C_{\rm Ham}$\ as the first class function (\ref{eq:C_ham}) that
remains after the gauge conditions have been introduced. It must
therefore be a combination of the original constraint functions as in
equation~(\ref{eq:C_ham}) that in addition satisfies
\[
\{ C_{\rm Ham}, \phi_i \} \approx 0, \quad i=1, 2, 3
\]
where the symbol `$\approx$' denotes equality on the surface defined
by imposing both the constraints and the gauge conditions. A simple
set of such conditions that was also used in Ref.~\cite{EffCons} to recover
the deparameterized dynamics of a Newtonian particle is provided by
\begin{eqnarray}
\phi_1 &=& (\Delta t)^2 = 0 \nonumber \\
\phi_2 &=& \Delta(tq) = 0 \nonumber \\
\phi_3 &=& \Delta(tp) = 0 \label{eq:standard_gauge}\,.
\end{eqnarray}
(Again, $(\Delta t)^2=0$ is consistent with the uncertainty relation
since $\Delta(tp_t)=-i\hbar/2$ from $C_{5\pm}$.)

Let $\Sigma_{\pm}$\ be the surfaces defined by simultaneously
imposing the constraints $\{C_{i\pm}\}$\ and the above gauge
conditions. These are coordinatized by the physical variables---$q$,
$p$, $(\Delta q)^2$, $\Delta(qp)$, $(\Delta p)^2$---and the one
remaining gauge degree of freedom---$t$. It is straightforward to
verify that the variables $(\Delta t)^2$, $\Delta(tq)$ and
$\Delta(tp)$\ generate a Poisson ideal of the algebra of physical
and gauge variables (i.e. the variables that do not involve
$\mathbf{p}_t$). That is, $\{\phi_i, X\}$\ is a sum of gauge
conditions with some coefficients. It follows that on $\Sigma_\pm$\
the gauge-fixing conditions have a trivial Poisson algebra
$\{\phi_i, \phi_j\} \approx 0$\ and a vanishing Poisson bracket with
the remaining free variables. It is not difficult to see that
$C_{1\pm}$\ remains first class on $\Sigma_\pm$. Since $E$\ is a
function of the ``physical'' variables only,  $\{ E, \phi_i \}
\approx 0$ and so
\[
\{ \phi_i, C_{1\pm} \} = \{ \phi_i, p_t \pm E \} \approx 0\,.
\]
Furthermore, writing $C_{5\pm} = \Delta(tp_t) \pm \phi_3p/E
+ {\rm const.}$\ one can quickly establish that $C_{5\pm}$\ also
remains first class but has a vanishing Poisson flow on
$\Sigma_\pm$.

The remaining set of surface-defining conditions composed of
$C_{2\pm}$, $C_{3\pm}$, $C_{4\pm}$ and $\{\phi_i\}_{i=1, 2, 3}$\ is
second class for all admissible values of the physical variables.
This can be seen by relabeling the conditions as $\chi_1 =
C_{2\pm}$, $\chi_2 = C_{3\pm}$, $\chi_3 = C_{4\pm}$, $\chi_4 =
\phi_1$, $\chi_5 = \phi_2$, $\chi_6 = \phi_3$\ and looking at the
Poisson bracket matrix $\Delta_{ij}:= \{ \chi_i, \chi_j\}$. On
$\Sigma_{\pm}$\ the components of the matrix are
\begin{equation*}
\mathbf{\Delta} \approx \left( \begin{array}{cccccc} 0 & 0 & 0 &
2i\hbar & \frac{\pm p}{E}(i\hbar + 2\Delta(qp)) & \frac{\pm 2p}{E}
(\Delta p)^2 \\ 0 & 0 & 0 & 0 & \frac{1}{2}i\hbar- \Delta(qp) &
-(\Delta p)^2 \\ 0 & 0 &
0 & 0 & -(\Delta q)^2 & -\frac{1}{2} i \hbar - \Delta(qp) \\
-2i\hbar & 0 & 0 & 0 & 0 & 0 \\ \frac{\mp p}{E} (i\hbar +
2\Delta(qp)) & \Delta(qp) - \frac{1}{2}i\hbar  & (\Delta q)^2 & 0 & 0 & 0 \\
\frac{\mp 2p}{E}(\Delta p)^2 & (\Delta p)^2 & \frac{1}{2} i\hbar +
\Delta(qp) & 0 & 0 & 0
\end{array} \right)\,.
\end{equation*}
Calculating the determinant one obtains the same result for both
choices of the sign
\[
\det[\mathbf{\Delta}] \approx -4\hbar^2 \left[ \frac{\hbar^4}{16}
+\left( \Delta(qp) \right)^4 + 2 \left( (\Delta p)^2(\Delta q)^2 -
\frac{\hbar^2}{4} \right) \left( (\Delta p)^2 (\Delta q)^2 - \left(
\Delta(qp) \right)^2 \right) \right]\,.
\]
The determinant is non-zero in the region where reality, positivity
and uncertainty conditions are imposed on the state---that
is, if one demands
\begin{eqnarray*}
&& q, p, (\Delta q)^2, \Delta(qp), (\Delta p)^2 \in \mathds{R} \\
&& (\Delta p)^2, (\Delta q)^2 \geq 0 \\ && (\Delta p)^2 (\Delta q)^2
- \left( \Delta(qp) \right)^2 \geq \frac{\hbar^2}{4}\,.
\end{eqnarray*}
With these conditions in place, the sum of the terms inside the
square bracket in the expression for the determinant is strictly
positive, which means that the determinant itself is strictly
negative.

There is one important check that one needs to perform. The
introduction of $\phi_i = 0$, ${i=1,2,3}$\ makes the surfaces
$\Sigma_{\pm}$\ a mixture of first and second class and one is
required to adjust the Poisson structure of the functions
parameterizing the surfaces through the use of the Dirac bracket.
Before we can identify $q$, $p$, $(\Delta q)^2$, $\Delta(qp)$,
$(\Delta p)^2$\ as the expectation values and moments of a
\emph{physical} canonical pair of operators, we need to verify that
their Dirac brackets on $\Sigma_{\pm}$\ are identical to the Poisson
brackets one would obtain for the quantum variables associated with
a single canonical pair. The bracket may be computed as follows
\begin{equation}
\left\{ f, g \right\}_{\rm Dirac}:=\left\{ f, g \right\} - \left\{
f, \chi_i \right\} (\Delta^{-1})^{ij} \left\{ \chi_j, g \right\}\,.
\end{equation}
Using the fact that $\Delta_{ij}$\ (and hence also
$(\Delta^{-1})^{ij}$) is off-block-diagonal and that the physical
variables have vanishing Poisson brackets with the gauge fixing
conditions, one can easily verify that the second term in the above
definition vanishes for the brackets between $q$, $p$, $(\Delta
q)^2$, $\Delta(qp)$, $(\Delta p)^2$\ (as well as $t$), and therefore
their Poisson structure is unchanged as required by our
interpretation. These variables are the remaining physical
quantities up to second order on the reduced quantum phase space.

To summarize, we impose the gauge-fixing conditions $\phi_i = 0$,
${i=1,2,3}$, interpret $q$, $p$, $(\Delta q)^2$, $\Delta(qp)$,
$(\Delta p)^2$ as the physical expectation values and moments and as
a result demand reality, positivity and quantum uncertainty. With
all of these conditions taken together, $\phi_i = 0$, ${i=1,2,3}$\
restrict the gauge freedom up to the orbits generated by $C_{1\pm}$\
(recall that $C_{5\pm}$\ generates no flow on $\Sigma_{\pm}$). This
means that the time evolution is given by
\[
C_{\rm Ham} = \mu_1 C_{1\pm}\,.
\]
Finally, we fix the remaining Lagrange multiplier $\mu_{1}$\ by
demanding $t$---the last remaining gauge variable---to be the time
parameter. That is, we demand that $\{t, C_{\rm Ham}\} \approx 1$,
which leaves us with
\[
C_{\rm Ham} =  C_{1\pm} = p_t \pm E\,.
\]

Taking the non-relativistic limit of $E$\ we recover the results for
a deparameterized free Newtonian particle. Specifically, if we formally
take $p^2/m^2$\ to be of order $\delta$, it follows that in
a semiclassical state $(\Delta p)^2/m^2$\ is of order $\hbar
\delta$. We expand the expression for $E$\ to the leading order in
$\delta$:
\begin{align*}
E =& \frac{m}{\sqrt{2}} \left( 1 + \frac{p^2 + (\Delta p)^2}{m^2}
\right)^{\frac{1}{2}} \left( 1 + \sqrt{ 1 - \frac{4p^2 (\Delta
p)^2}{(p^2 + m^2 + (\Delta p)^2)^2}} \right)^{\frac{1}{2}}
\\ =& \frac{m}{\sqrt{2}} \left( 1 + \frac{p^2 + (\Delta p)^2}{2m^2}
+ O(\delta^2) \right) \left( 2 + O(\hbar \delta^2)
\right)^{\frac{1}{2}} \\ =& m + \frac{p^2 + (\Delta p)^2}{2m} +
O(\delta^2)\,.
\end{align*}

\subsection{Comparison with the Klein-Gordon solution}
\label{sec:klein_gordon}

The standard positive frequency solutions to the Klein-Gordon
equation (see for example~\cite{LocalQuant}) form a Hilbert space of
momentum-space wave-functions square integrable with respect to the
Lorentz-invariant measure:
\[
\mathcal{H} = L^2 \left( \mathds{R}, \frac{\mathrm{d}k}{2\epsilon_k} \right),
\quad {\rm where} \quad \epsilon_k = \sqrt{k^2+m^2}\,.
\]
The system can be described through a canonical pair of observables,
represented on $\mathcal{H}$ as
\[
\mathbf{p} = k \quad {\rm and} \quad \mathbf{q} = i\hbar \left(
\frac{\partial}{\partial k} + \epsilon_k \left(
\frac{\partial}{\partial k} \frac{1}{2 \epsilon_k} \right) \right)\,.
\]
The time evolution is generated by the Hamiltonian $\mathbf{H} =
\left( \mathbf{p}^2 + m^2\mathds{1} \right)^{\frac{1}{2}}$. One can
evaluate the evolution equations for the expectation values of the
observables using Ehrenfest's theorem
\[
\frac{\mathrm{d}}{\mathrm{d}t} \langle \mathbf{O} \rangle = \frac{1}{i\hbar}
\left\langle \left[ \mathbf{O}, \mathbf{H} \right] \right\rangle
+\frac{\partial\langle\mathbf{O}\rangle}{\partial t}\,.
\]
In our formalism, the right-hand side is equivalent to the quantum
Poisson bracket between the expectation values, thus
\begin{align*}
\frac{\mathrm{d}}{\mathrm{d}t} \langle \mathbf{O} \rangle &= \left\{
\langle \mathbf{O} \rangle, \langle \mathbf{H} \rangle \right\} +
\frac{\partial\langle\mathbf{O}\rangle}{\partial t} = \left\{
\langle \mathbf{O} \rangle, \langle  \left( \mathbf{p}^2 +
m^2\mathds{1} \right)^{\frac{1}{2}} \rangle \right\} +
\frac{\partial\langle\mathbf{O}\rangle}{\partial t}\,.
\end{align*}
We recall that our procedure at order $\hbar$\ together with the
gauge fixing conditions for the positive frequency solutions
resulted in
\begin{align*}
\frac{\mathrm{d}}{\mathrm{d}t} \langle \mathbf{O} \rangle = \left\{
\langle \mathbf{O} \rangle, p_t + E \right\} = \left\{ \langle
\mathbf{O} \rangle, E \right\}
+\frac{\partial\langle\mathbf{O}\rangle}{\partial t}\,.
\end{align*}
In order to see whether the methods agree, we only need to compare
$\langle \left( \mathbf{p}^2 + m^2\mathds{1} \right)^{\frac{1}{2}}
\rangle$\ and $E$\ to order $\hbar$. To verify this explicitly we
expand the operator in terms of its moments, assuming the
expectation value to be taken in a semiclassical state:
\begin{align*}
\left\langle (\mathbf{p}^2 + m^2 \mathds{1})^{\frac{1}{2}}
\right\rangle =& \left\langle(p^2 + m^2)^{\frac{1}{2}} +
\frac{p}{(p^2 + m^2)^{\frac{1}{2}}} (\mathbf{p} - p) +
\frac{m^2}{2(p^2 + m^2)^{\frac{3}{2}}}(\mathbf{p} -
p)^2\right\rangle +
({\rm higher \ moments})\\
=& \sqrt{p^2 + m^2} \left( 1 + \frac{m^2 (\Delta p)^2}{ 2( p^2 +
m^2)^2} \right) + O(\hbar^{\frac{3}{2}})\,.
\end{align*}
For comparison, we expand $E$\ in powers of $(\Delta p)^2$, which we
assume to be of order $\hbar$.
\begin{align*}
E &= \frac{1}{\sqrt{2}} \sqrt{ p^2 + m^2} \left( 1 + \frac{(\Delta
p)^2}{p^2 + m^2} + 1 + \frac{(m^2 - p^2) (\Delta p)^2}{ ( p^2 +
m^2)^2} + O\left((\Delta p)^4\right) \right)^{\frac{1}{2}} \\
&=\sqrt{p^2 + m^2} \sqrt{ 1 + \frac{m^2 (\Delta p)^2}{( p^2 + m^2)^2}
+ O\left((\Delta p)^4\right)}\\ &= \sqrt{p^2 + m^2} \left( 1 +
\frac{m^2 (\Delta p)^2}{ 2( p^2 + m^2)^2} \right) + O\left((\Delta
p)^4\right)\,.
\end{align*}
Thus, up to the terms of order $\hbar$\ the two results agree.

Unlike the exact Klein-Gordon solution, our approach avoids explicit
reference to a representation. The action of the Lorentz group on
our variables can be understood through its action on the algebra of
observables. In particular, we assume that the pairs $\left(
\mathbf{p}_t, \mathbf{p} \right)$\ and $\left( \mathbf{t}
,\mathbf{q} \right)$\ transform as components of a contravariant and
a covariant vector respectively. Looking at the truncated system of
constraints~(\ref{eq:2_ord_constr_m}) under a Lorentz transformation
one finds that $C$\ remains invariant, while the pairs $\left(
C_{p_t}, C_p \right)$\ and $\left( C_t, C_q \right)$\ themselves
transform as components of a contravariant and a covariant vector
respectively, so that the whole truncated system of constraints is
preserved.

\subsection{Free massless particle}

For a massless particle, the constraint operator takes the form
\[
\mathbf{C} = \mathbf{p}_t^2 - \mathbf{p}^2\,.
\]
To second order in moments, the constraint functions produced remain
as in equation~(\ref{eq:2_ord_constr_m}), except for $C$, which now
reads
\[
C = \langle \mathbf{C} \rangle = p_t^2 - p^2 + (\Delta p_t)^2 -
(\Delta p)^2 = 0\,.
\]
The disappearance of a constant term from $C$\ does not affect the
Poisson algebra of the constraints, so that the table of
Section~\ref{sec:massive_particle_constraints} still applies. The
solution to the constraints, however takes on a simpler form:
following the same steps as previously and eliminating $(\Delta
p_t)^2$\ in a way compatible with the semiclassical approximation we
obtain
\[
(\Delta p_t)^2 = (\Delta p)^2\,.
\]
Together with $C=0$\ this implies
\[
p_t^2 = p^2\,.
\]
As we see, the classical constraint is satisfied \emph{exactly} by
the expectation values. We solve this via
\[
p_t = \pm |p|\,.
\]
There are two related reasons for taking the absolute value of $p$\
in the above solution. Firstly, to emphasize the sign of the energy.
Secondly, to match the limit as $m$\ is set to zero of the solution
obtained in Section~\ref{sec:massive_particle_constraints}. The full
solutions read
\begin{eqnarray}
p_t &=& \pm |p| \nonumber \\ \Delta(tp_t) &=& \pm \frac{p}{|p|}
\Delta(t p)
- \frac{i\hbar}{2} \nonumber \\ (\Delta p_t)^2 &=& (\Delta p)^2 \nonumber \\
\Delta(p_tq) &=& \pm \frac{p}{|p|}\left( \Delta(qp) +
\frac{i\hbar}{2} \right) \nonumber \\ \Delta(p_tp) &=& \pm
\frac{p}{|p|} (\Delta p)^2 \label{eq:2ord_sol2}\,.
\end{eqnarray}
The steps of Section~\ref{sec:massive_particle_gauge} can be
repeated exactly for the $m=0$\ case with $|p|$\ playing the role of
$E$. With the gauges fixed in an identical way, this results in
evolution on $q$, $p$, $(\Delta q)^2$, $\Delta(qp)$, $(\Delta p)^2$\
generated by the constraint
\begin{equation} \label{CHammassless}
C_{\rm Ham} = p_t \pm |p|\,.
\end{equation}
The implications will be discussed further in the conclusions.

\section{Relativistic particle in a potential}
\label{sec:particle_in_pot}

In this section we consider the consequences of adding a potential
term to the quantum constraint. We consider a quadratic
time-independent potential in Section~\ref{sec:space_pot} followed
by a homogeneous time-dependent potential in
Section~\ref{sec:time_pot}. The systems considered in this section
have the same kinematical degrees of freedom as the free
relativistic particle; however, the additional terms in the
constraint element break Lorentz invariance. On the other hand,
certain structural properties of the constraint element remain very
similar to the free particle case, which makes extension of the
calculations performed in Section~\ref{sec:free_particle} fairly
simple. As we will see, the constraints are still straightforward to
solve, but their Poisson algebra is only approximately closed
and in the case of the time-dependent potential, the gauge analysis
requires more subtlety. These examples show that the effective
procedure used here is feasibly applicable to a wider range of
models than the existing explicit constructions of a physical inner
product.

\subsection{Quadratic potential}\label{sec:space_pot}
A relativistic particle in a quadratic potential is subject to the
constraint
\[
\mathbf{C} = \mathbf{p}_t^2 - \mathbf{p}^2 - \mathbf{q}^2 - m^2
\mathds{1}\,.
\]
(A coupling constant in the potential could be absorbed by rescaling.)
This gives rise to the following set of constraint functions truncated
at second order
\begin{align}
C& = p_t^2 - p^2 - q^2 - m^2 + (\Delta p_t)^2 - (\Delta p)^2 -
(\Delta q)^2 = 0 \nonumber \\ C_{t}& = 2p_t \Delta(t p_t) + i\hbar
p_t - 2p \Delta(t p) - 2q \Delta(tq) = 0 \nonumber \\ C_{p_t}& =
2p_t (\Delta p_t)^2 - 2p \Delta(p_tp) -2q\Delta(p_tq) = 0 \nonumber
\\ C_q& = 2p_t \Delta(p_tq) - 2p \Delta(qp) - i\hbar p
- 2q(\Delta q)^2 = 0 \nonumber
\\ C_p& = 2p_t \Delta(p_tp) - 2p(\Delta p)^2 -2q \Delta(qp) +
i\hbar q = 0\,. \label{eq:2_ord_constr_pot}
\end{align}
The above system of constraints is first class only to order $\hbar$
as can be seen from their Poisson algebra in Table~\ref{tab:pb_pot}.

\begin{table*}
\caption{Poisson algebra of constraints for the particle in a
quadratic potential. First terms in the bracket are labeled by rows,
second terms are labeled by columns. } \centering
\begin{tabular}{|c||c|c|c|c|c|} \hline & ${\scriptstyle C}$ & ${\scriptstyle C_{t}}$ &
${\scriptstyle C_{p_t}}$ & ${\scriptstyle C_q}$ & ${\scriptstyle C_p}$ \\
\hline \hline

${\scriptstyle C}$ & ${\scriptstyle 0}$ & ${\scriptstyle -2C_{p_t}}$
& ${\scriptstyle 0}$ & ${\scriptstyle 2C_p}$ & ${\scriptstyle
-2C_q}$
\\ \hline

& & & ${\scriptstyle 4p_tC_{p_t} - 2pC_p - 2qC_q}$ & ${\scriptstyle
2p_tC_q + 2pC_{t}}$ & ${\scriptstyle 2p_tC_p + 2qC_{t}}$ \\
${\scriptstyle C_{t}}$ & ${\scriptstyle 2C_{p_t}}$ & ${\scriptstyle
0}$ & ${\scriptstyle +4\Delta(p_tp) \Delta(tq)}$ & ${\scriptstyle
+4\Delta(tq) \left( \Delta(qp) + \frac{i\hbar}{2} \right)}$ &
${\scriptstyle +4\Delta(tq) (\Delta p)^2 }$ \\ & & & ${\scriptstyle
-4\Delta(p_tq) \Delta(tp)}$ & ${\scriptstyle +4 (\Delta q)^2
\Delta(tp)}$ &
${\scriptstyle -4\Delta(tp) \left(\Delta(qp) - \frac{i\hbar}{2} \right)}$ \\
\hline

& & ${\scriptstyle -4p_tC_{p_t} + 2pC_p + 2qC_q}$ & & ${\scriptstyle
2pC_{p_t}}$ & ${\scriptstyle -2q C_{p_t}}$ \\ ${\scriptstyle
C_{p_t}}$ & ${\scriptstyle 0}$ & ${\scriptstyle -4\Delta(p_tp)
\Delta(tq)}$ & ${\scriptstyle 0}$ & ${\scriptstyle -4(\Delta q)^2
\Delta(p_tp)}$ & ${\scriptstyle +4 (\Delta p)^2 \Delta(p_tq)}$ \\ &
& ${\scriptstyle +4\Delta(p_tq) \Delta(tp)}$ & & ${\scriptstyle
-\Delta(p_tq) \left(\Delta(qp) - \frac{i\hbar}{2} \right)}$ &
${\scriptstyle +4\Delta(p_tp) \left(\Delta(qp) - \frac{i\hbar}{2} \right)}$ \\
\hline

& & ${\scriptstyle -2p_tC_q - 2pC_{t}}$ & ${\scriptstyle
-2pC_{p_t}}$ & & ${\scriptstyle 2p_tC_{p_t} - 4pC_p - 4qC_q}$ \\
${\scriptstyle C_q}$ & ${\scriptstyle -2C_p}$ & ${\scriptstyle
-4\Delta(tq) \left(\Delta(qp) + \frac{i\hbar}{2} \right)}$ &
${\scriptstyle + 4(\Delta q)^2 \Delta(p_tp)}$ & ${\scriptstyle 0}$ &
${\scriptstyle +4
\left( (\Delta q)^2 (\Delta p)^2 - \frac{\hbar^2}{4} \right) }$ \\
& & ${\scriptstyle -4 \Delta(tp) (\Delta q)^2 }$ & ${\scriptstyle +4
\Delta(p_tq) \left(\Delta(qp) - \frac{i\hbar}{2}
\right)}$ & & ${\scriptstyle  -4 \left( \Delta(qp) \right)^2}$\\
\hline

& & ${\scriptstyle -2p_tC_p + 2qC_{t}}$ & ${\scriptstyle 2qC_{p_t}}$
& ${\scriptstyle 4pC_p + 4qC_q - 2p_tC_{p_t}}$ & \\ ${\scriptstyle
C_p}$ & ${\scriptstyle 2C_q}$ & ${\scriptstyle -4 \Delta(tq) (\Delta
p)^2}$ & ${\scriptstyle -4(\Delta p)^2 \Delta(p_tq)}$ &
${\scriptstyle -4 \left( (\Delta q)^2 (\Delta p)^2 -
\frac{\hbar^2}{4} \right) }$ & ${\scriptstyle 0}$ \\ & &
${\scriptstyle +4\Delta(tp) \left(\Delta(qp) - \frac{i\hbar}{2}
\right)}$ & ${\scriptstyle + 4\Delta(p_tp) \left(\Delta(qp) -
\frac{i\hbar}{2} \right)}$ & ${\scriptstyle  -4 \left( \Delta(qp)
\right)^2}$ & \\ \hline
\end{tabular}
\label{tab:pb_pot}
\end{table*}

The system of constraints may be solved following the same steps
that have been employed to solve the constraints for the free
particle. We use $C$, $C_{q}$, and $C_p$\ to eliminate $p_t^2$,
$\Delta(p_tq)$\ and $\Delta(p_tp)$\ respectively. Substituted into
$C_{p_t}$\ this yields a quadratic equation in $(\Delta p_t)^2$:
\begin{align*}
0 =& \left( (\Delta p_t)^2 \right)^2 - \left( p^2 + q^2 + m^2 +
(\Delta p)^2 + (\Delta q)^2 \right) (\Delta p_t)^2 + \left(
p^2(\Delta p)^2 + 2qp \Delta(qp) + q^2 (\Delta q)^2 \right)\,.
\end{align*}
The solution compatible with the semiclassical approximation has the
form
\begin{eqnarray}
p_t &=& \pm E \nonumber \\
\Delta(tp_t) &=& \frac{\pm 1}{E} \left( p \Delta(tp) + q \Delta(tq)
\right) - \frac{i\hbar}{2} \nonumber \\ (\Delta p_t)^2 &=& p^2 + q^2
+ m^2 + (\Delta p)^2 + (\Delta q)^2 - E^2 \nonumber \\
\Delta(p_tq) &=& \frac{\pm 1}{E} \left( p
\Delta(qp) + \frac{i\hbar}{2}p + q(\Delta q)^2 \right) \nonumber \\
\Delta(p_tp) &=& \frac{\pm 1}{E} \left( p(\Delta p)^2 + q\Delta(qp)
- \frac{i\hbar}{2}q \right) \label{eq:2nd_ord_sol_pot}
\end{eqnarray}
where
\begin{eqnarray} \label{CHam}
E &=& \frac{1}{\sqrt{2}} \Biggl[ p^2 + q^2 + m^2 +(\Delta p)^2 +
(\Delta q)^2  \\ && + \biggl( (p^2 + q^2 + m^2 + (\Delta
p)^2 + (\Delta q)^2 )^2   - 4 \left( p^2(\Delta p)^2 + 2qp
\Delta(qp) + q^2(\Delta q)^2 \right) \biggr)^{\frac{1}{2}}
\Biggr]^{\frac{1}{2}}\,.\nonumber
\end{eqnarray}
We note that as $C = p_t^2 - E^2$\ and $p_t$\ are both exactly first
class, $E$\ must also be first class, since it does not vanish on
the constraint surface.

If one applies the semiclassical approximation once again to drop
the terms of orders higher than $\hbar$, the constraint system may
be treated as first class. The linearized versions of the
constraints take the form
\begin{eqnarray}
C_{1\pm} &=&  p_t \pm E \nonumber \\ C_{2\pm} &=& (\Delta
p_t)^2 - p^2 - q^2 - m^2 - (\Delta p)^2 - (\Delta q)^2 + E^2 \nonumber \\
C_{3\pm} &=& \Delta(p_tp) \pm \frac{1}{E} \left( p(\Delta p)^2 + q
\Delta(qp) - q \frac{i\hbar}{2} \right) \nonumber \\ C_{4\pm} &=&
\Delta(p_tq) \pm \frac{1}{E} \left( p \Delta(qp) + p
\frac{i\hbar}{2} + q (\Delta q)^2 \right) \nonumber \\ C_{5\pm} &=&
\Delta(tp_t) \pm \frac{1}{E} \left( p \Delta(t p) + q \Delta(tq)
\right) + \frac{i\hbar}{2}\,.
\end{eqnarray}
One can follow the process outlined in the
Section~\ref{sec:massive_particle_gauge} and impose the set of
gauge-fixing conditions~(\ref{eq:standard_gauge}). Once again
$C_{1\pm}$\ is first-class on the gauge-fixed surfaces
$\Sigma_{\pm}$\ and writing $C_{5\pm} = \Delta(tp_t) \pm \frac{p}{E}
\phi_3 \pm \frac{q}{E} \phi_2$\ it is not difficult to see that,
once again, $C_{5\pm}$\ remains first-class but has a vanishing flow
on $\Sigma_{\pm}$. We recall that the gauge conditions have a
vanishing flow on the remaining free variables; therefore only the
first term in the expressions for each of the constraints
$C_{i\pm}$\ above has a non-vanishing Poisson bracket with the
conditions $\phi_i$. As a result, the Poisson bracket matrix
$\mathbf{\Delta}$\ remains as in
Section~\ref{sec:massive_particle_gauge} up to entries of order
$\hbar$. Imposing reality, positivity and quantum uncertainty and
demanding $\{t, C_{\rm Ham} \} \approx 1$\ we once again obtain
\[
C_{\rm Ham} = p_t \pm E\,.
\]
Directly expanding $E$\ in powers of $(\Delta q)^2$, $\Delta(qp)$\
and $(\Delta p)^2$\ we get the expression to order $\hbar$
\begin{multline}
E = \sqrt{p^2 + q ^2 + m^2}\Biggl[ 1 + \frac{ (q^2 +
m^2)(\Delta p )^2 - 2qp \Delta(qp) + (p^2 + m^2) (\Delta q)^2}{2(p^2
+ q ^2 + m^2)^2} \Biggr] \\ + O\left((\Delta p)^4\right) +
O\left((\Delta q)^4\right) + O\left((\Delta (qp))^2\right) \,.
\label{eq:E_space_pot}
\end{multline}

The specific constraint considered in this section can be implemented
quite completely at the level of physical states, although specifics
of the dynamics are more difficult to extract than with effective
methods. Below we briefly describe the solution and compare it to our
effective treatment. The algebra elements may be represented
kinematically as differential operators on the space of
square-integrable wave-functions in two variables $x_0$\ and $x_1$\ in
the usual way
\[
\mathbf{t} = x_0,\quad \mathbf{p}_t = \frac{\hbar}{i}
\frac{\partial}{\partial x_0}, \quad \mathbf{q} = x_1, \quad
\mathbf{p} = \frac{\hbar}{i} \frac{\partial}{\partial x_1}\,.
\]
The constraint operator splits into a sum of commuting, and
therefore simultaneously diagonalizable,
components:

\begin{itemize}

\item $\mathbf{p}_t^2 = -\hbar^2
\frac{\partial^2}{\partial x_0^2}$\ has infinite-norm eigenstates of
the form $\exp(\frac{ik}{\hbar}x_0)\phi(x_1)$, with eigenvalues
$k^2$.

\item $\mathbf{p}^2 + \mathbf{q}^2= -\hbar^2\frac{\partial^2}{\partial x_1^2} + x_1^2
= 2 \mathbf{H}_{\rm harm}$, where $\mathbf{H}_{\rm harm}$ is
precisely the standard Hamiltonian of the harmonic oscillator on
$x_1$ (with mass and frequency set to unity). This operator has
normalizable eigenstates of the form $\psi(x_0) \varphi_n(x_1)$,
where $\varphi_n(x_1)$ is the usual normalized $n$-th eigenstate of
the harmonic oscillator, the corresponding eigenvalues are
$2(n+\frac{1}{2})\hbar$.

\item Every wavefunction is an eigenstate of $\mathds{1}m^2$\
with the eigenvalue $m^2$.

\end{itemize}

The eigenfunctions of the constraint operator are $\Psi_{k, n}(x_0,
x_1) = \exp(\frac{ik}{\hbar}x_0)\varphi_n(x_1)$, the corresponding
eigenvalues are $\left( k^2 - 2(n+\frac{1}{2})\hbar - m^2 \right)$.
The space of solutions to the quantum constraint equation is
therefore spanned by the wavefunctions $\Psi_{k, n}$\ for which $k=
\pm \sqrt{2(n+\frac{1}{2})\hbar + m^2}$. These states are not
normalizable with respect to the square integration in both $x_0$\
and $x_1$, however they have unit norm when the integration is taken
with respect to $x_1$\ alone.

As is usually done for such systems we decompose the solution space
into two segments one belongs to the positive part, the other one to
the negative part of the spectrum of $\mathbf{p}_t$ (e.g. the
separation of positive and negative frequencies of the solutions to
the Klein-Gordon equation). The general element of the solution
space is a linear combination of either positive or negative
frequency null eigenfunctions of $\mathbf{C}$, denoted by $\Psi^+$\
and $\Psi^-$ respectively
\[
\Psi_{\rm phys}^{\pm} = \sum_{n=0}^{\infty} \alpha_n \exp \left(
\frac{\mp i x_0 \sqrt{2(n+\frac{1}{2})\hbar + m^2}}{\hbar} \right)
\varphi_n(x_1)\,.
\]
The separation into two components allows us to define a
positive-definite physical inner product on each one of them
individually
\begin{equation}\label{PhysInn}
\langle \Psi | \Phi \rangle_{\rm phys} := \int_{-\infty}^{\infty}
\overline{\Psi(x_0, x_1)} \Phi(x_0, x_1) {\rm d}x_1
\end{equation}
where the bar denotes a complex-conjugate and both states $\Psi$,
$\Phi$\ belong to the same component. On the space of solutions, the
above inner product is independent of the value taken by $x_0$.
Furthermore this inner product is consistent with interpreting
$\mathbf{t}$\ as time, since we can formally write $\langle \Psi |
\mathbf{t} |\Psi \rangle_{\rm phys} = x_0$.  (This equation requires
some care in its interpretation since $\mathbf{t}$ is not a physical
operator. But just viewing the integration on the right hand side of
(\ref{PhysInn}) easily allows us to introduce an operator $\mathbf{t}$
by multiplication. The dependence on $x_0$ of the result is then in
agreement with the fact that $\mathbf{t}$ is not a physical
observable.)  It is also consistent with the gauge choices of
equation~(\ref{eq:standard_gauge}), which is straightforward to verify
using the fact that for any operator $\mathbf{A}$, polynomial in
$\mathbf{t}, \mathbf{p}_t, \mathbf{q}, \mathbf{p}$
\[
\langle \Psi | \mathbf{tA} |\Psi \rangle_{\rm phys} = x_0 \langle
\Psi | \mathbf{A} |\Psi \rangle_{\rm phys}\,.
\]
The physical states may be interpreted as solutions of one of the
two Scr\"odinger equations
\[
-\frac{\hbar}{i} \frac{\rm d \ }{{\rm d} x_0} \Psi^{\pm} = \pm
\left( \mathbf{p}^2 + \mathbf{q}^2 + m^2\mathds{1}
\right)^{\frac{1}{2}} \Psi^{\pm}\,.
\]
That is, an ordinary quantum mechanical system with time evolution in
the variable $x_0$\ generated by the self-adjoint, positive
square-root Hamiltonian $\mathbf{H} = \left( \mathbf{p}^2 +
\mathbf{q}^2 + m^2\mathds{1}\right) ^{\frac{1}{2}}$, which is defined
through its action on the basis of eigenstates: $\mathbf{H}
\varphi_n(x_1) = \sqrt{2(n+\frac{1}{2})\hbar + m^2} \varphi_n(x_1)$.

To compare the physical states and the effective solutions, we proceed as we
have done before, in Section~\ref{sec:klein_gordon}. We expand the
expectation value of the square-root hamiltonian in a semiclassical
state
\begin{align*}
\langle \mathbf{H} \rangle =&  \left\langle \left\langle
\mathbf{p}^2 + \mathbf{q}^2 + m^2\mathds{1}
\right\rangle^{\frac{1}{2}} + \frac{\left(\mathbf{p}^2 +
\mathbf{q}^2 + m^2\mathds{1} \right) - \left\langle \mathbf{p}^2 +
\mathbf{q}^2 + m^2\mathds{1} \right\rangle}{2 \left\langle
\mathbf{p}^2 + \mathbf{q}^2 + m^2\mathds{1}
\right\rangle^{\frac{1}{2}}}\right.\\
&\left. - \frac{\left( \left(\mathbf{p}^2 +
\mathbf{q}^2 + m^2\mathds{1} \right) - \left\langle \mathbf{p}^2 +
\mathbf{q}^2 + m^2\mathds{1} \right\rangle\right)^2}{8 \left\langle
\mathbf{p}^2 + \mathbf{q}^2 + m^2\mathds{1}
\right\rangle^{\frac{3}{2}}} \right\rangle +({\rm higher \ moments})\,.
\end{align*}
Proceeding with the above expansion and keeping only the terms up to
order $\hbar$ one obtains the expression that is identical to the one
for $E$\ in equation~(\ref{eq:E_space_pot}). Thus, to leading order in
the semiclassical regime, the effective solution to the constraint is
consistent with physical state evolution, and the gauge choice of
equation~(\ref{eq:standard_gauge}) is consistent with the physical
inner product defined above together with the interpretation of
$\langle \mathbf{t} \rangle$\ as measuring the physical time. For a
direct comparison between fully quantum and effective time evolutions
for a specific semiclassical state of this system see
App.~\ref{AppII}.

If we replace $\left( \mathbf{p}^2 + \mathbf{q}^2 +
m^2\mathds{1}\right)$\ in the constraint by any positive operator
$f(\mathbf{q},\mathbf{p})$ analytic in $\mathbf{q}$\ and $\mathbf{p}$,
physical states can in principle be found in a similar way. One could
find the spectrum of $f(\mathbf{q},\mathbf{p})$\ and construct the
solutions out of its simultaneous eigenfunctions with
$\mathbf{p}_t^2$. Finding the spectrum of a given operator is in
general a complicated task.  Further, we were helped in this example
by the fact that the spectrum of $(\mathbf{p}^2 + \mathbf{q}^2 +
m^2\mathds{1})$\ is discrete and the eigenfunctions are normalizable
with respect to square integration over $x_1$\ alone. Defining the
physical inner product is more difficult if parts of the spectrum of
$f(\mathbf{q},\mathbf{p})$\ are continuous. Finally, determining
suitable coherent states for semiclassical purposes, as done for this
model in App.~\ref{AppII}, can be a difficult task. The leading order
effective solution, on the other hand, can be obtained in much the
same way as was done for the above example, without explicit in-depth
knowledge of the spectrum of $\mathbf{q}$\ or the exact form of its
eigenstates.

\subsection{Time-dependent potential}
\label{sec:time_pot}

As mentioned in the introduction, time-dependent potentials are of
interest in quantum cosmology. Another example where time-dependent
terms arise is a relativistic particle moving on a non-static curved
background space-time. In such cases, our methods can be used as well,
but additional subtleties do arise. Adding a ``potential''
$V(t)=\lambda t$ to the classical constraint gives the second order
quantum constraints
\begin{align}
C& = p_t^2 - p^2 - m^2 + (\Delta p_t)^2 - (\Delta p)^2+\lambda t = 0
\nonumber \\ C_{t}& = 2p_t \Delta(t p_t) + i\hbar p_t - 2p \Delta(t
p)+\lambda (\Delta t)^2 = 0 \nonumber \\ C_{p_t}& = 2p_t (\Delta
p_t)^2 - 2p \Delta(p_tp) + \lambda\Delta(p_tt) -
\frac{1}{2}i\lambda\hbar = 0 \nonumber
\\ C_q& = 2p_t \Delta(p_tq) - 2p \Delta(qp) - i\hbar
p+\lambda\Delta(qt) = 0 \nonumber \\ C_p& = 2p_t \Delta(p_tp) -
2p(\Delta p)^2+\lambda\Delta(tp) = 0\,.
\label{eq:2_ord_constr_time_pot}
\end{align}
These constraints, once again, form a closed Poisson algebra only up
to order $\hbar$. We proceed to solve the above set of polynomial
equations explicitly---$p_tC_{p_t}=0$\ implies
\[
p_t^2 (\Delta p_t)^2 - p_tp \Delta(p_t p) + \frac{1}{2}\lambda p_t
\Delta(t p_t) - \frac{i\hbar}{4}\lambda p_t = 0\,.
\]
Using $pC_p=0$ and $\lambda C_t = 0$ to eliminate $p_tp\Delta(p_t
p)$\ and $\lambda p_t \Delta(tp_t)$\ respectively we obtain
\[
p_t^2(\Delta p_t)^2 - p^2 (\Delta p)^2 + \lambda p \Delta(tp) -
\frac{1}{4} \lambda^2 (\Delta t)^2 - \frac{i\hbar}{2} \lambda p_t =
0 \,.
\]
Finally, we eliminate $(\Delta p_t)^2$\ using $C=0$\ to obtain a
quartic equation in $p_t$
\begin{align*}
0 =& p_t^4 - \left( p^2 + m^2 - \lambda t + (\Delta p)^2 \right)
p_t^2 + \frac{i\hbar}{2} \lambda p_t + \left( p^2 (\Delta p)^2 +
\frac{1}{4} \lambda^2 (\Delta t)^2 - \lambda p \Delta(tp) \right)\,.
\end{align*}
The exact solutions to the above quartic equation are of course
readily available, though they involve long algebraic expressions
and are not particularly illuminating. Furthermore, due to the
linear term in $p_t$\ in the equation, the gauge choices we have
employed previously together with the reality conditions would lead
us to conclude that $p_t$\ is complex. A more subtle gauge analysis
is required to solve the constraint without further approximations.
For instance, to respect reality conditions one would use a gauge
relating moments to the expectation values, for example by making
$\Delta(tp)$ dependent on $p_t$.

However, assuming the potential changes very slowly allows one to
move forward with the standard gauge choice. Treating $\lambda$ as a
second small parameter in addition to $\hbar^{1/2}$, such that
$\lambda\hbar$ is of higher than second order and discarding the
terms of order higher than $\hbar$\ we are left with
\[
p_t^4 - \left( p^2 + m^2 - \lambda t + (\Delta p)^2 \right) p_t^2 +
p^2 (\Delta p)^2 = 0\,.
\]
This equation could also be obtained by directly dropping products
of $\lambda$\ and second order moments or $\hbar$\ in the
expressions for the constraint
functions~(\ref{eq:2_ord_constr_time_pot}) and solving them. It is a
quadratic equation in $p_t^2$, with solutions that are much easier
to interpret. Compatibility with the semiclassical approximation
once again selects for us a set of solutions that have a very
similar form to those in~(\ref{eq:2nd_ord_sol_m}),
~(\ref{eq:2ord_sol2}), and~(\ref{eq:2nd_ord_sol_pot}), with
\[
E = \frac{1}{\sqrt{2}} \Biggl[ p^2 + m^2 - \lambda t + (\Delta p)^2
+ \left( (p^2 + m^2 - \lambda t + (\Delta p)^2)^2 - 4p^2(\Delta p)^2
\right)^{\frac{1}{2}} \Biggr]^{\frac{1}{2}}\,.
\]
One can then repeat the gauge-fixing procedure we have previously
employed and recover $C_{\rm Ham} = p_t \pm E$, for a slowly varying
potential in a semiclassical state. Thus, to the semiclassical order
considered, the system behaves as a non-relativistic quantum particle
in one dimension subject to a time-dependent hamiltonian $\mathbf{H} =
\left(\mathbf{p}^2 + m^2 \mathds{1} - \lambda t
\right)^{\frac{1}{2}}$.

In fact, a more general ``slowly varying'' potential may be treated
to order $\hbar$\ in an analogous manner. We assume that the
potential has the form $V(\mathbf{t}) = V(0)\mathds{1} + \lambda
\tilde{V}(\mathbf{t})$, where $\lambda$\ is small in the sense
discussed earlier, and $\tilde{V}(\mathbf{t})$\ is a polynomial in
$\mathbf{t}$. This implies, in particular, that $\langle \lambda
\tilde{V}(\mathbf{t}) \rangle = \lambda \tilde{V}(t) +
O(\hbar^{\frac{3}{2}})$. We absorb the constant part of the
potential into the definition of $m$ and the constraint functions to
order $\hbar$\ look exactly as they do for a free relativistic
particle~(\ref{eq:2_ord_constr_m}), except for $C$, which acquires
an extra term
\[
C = p_t^2 - p^2 - m^2 + (\Delta p_t)^2 - (\Delta p)^2+\lambda
\tilde{V}(t) = 0\,.
\]
The resulting system of constraint functions may be solved,
gauge-fixed and interpreted directly following the methods employed
throughout Sections~\ref{sec:free_particle}
and~\ref{sec:particle_in_pot}. This demonstrates the flexibility of
the constructions, confirming the methods of
\cite{BouncePot,QuantumBounce,BounceSqueezed}, where also slowly
varying potentials were assumed. In contrast to this earlier work,
the general methods presented here are in principle applicable to
arbitrary potentials, but the gauge fixing would have to be
considered in each case in detail. This provides access to questions
about the role of time when potentials forbid a global monotonic
internal time $t$, resulting in a new perspective to be followed
elsewhere.

\section{Conclusion}

One of the main hurdles for quantum gravity is the physical Hilbert
space issue. At least for semiclassical questions, technical and
conceptual difficulties can be circumvented by using expectation
values and moments directly rather than states. Other advantages of
this method are that the specification of semiclassical regimes is
easier via moments (while semiclassical wave functions are often
difficult to formulate, even simple-looking Gaussian ones not always
being semiclassical at all in some models of quantum cosmology as
pointed out in \cite{NonExpLQC}) and that density states are
automatically included.

We have extended the methods for effective constraints of
\cite{EffCons} to relativistic systems, clarifying several
physically relevant issues of effective equations:
\begin{itemize}
\item Square root effective Hamiltonians, which so far were strictly
justified only for time-independent potentials, are valid even in
the time dependent case provided the potential varies slowly in
time. No extra conditions on the dynamics are implied by the
positivity conditions.
\item Quantum back-reaction follows reliably from square-root
Hamiltonians. Our examples of relativistic systems have shown three
different cases:
\begin{itemize}
\item Massless particles do not contain moments in their reduced
Hamiltonian (\ref{CHammassless}) and thus are not subject to quantum
back-reaction.
\item Free massive particles do have quantum back-reaction from
moments of all orders, which is initially unexpected since the
quantum constraint has only a linear term of $(\Delta p)^2$ with no
coupling to the expectation values. (Our formulas, done here only to
second order, do not show this explicitly.)
\item Particles subject to a $q$-dependent potential receive quantum
back-reaction from the covariance of their wave function as seen in
(\ref{CHam}). This is also unexpected since the expectation value of
the constraint does not contain mixed moments. As in the case of
massive particles, the unexpected results are explained by the
presence of higher order constraints.
\end{itemize}
\item Higher order constraints, which are crucial for the effective
procedure, also affect the amount of spreading of wave functions, or
the time dependence of moments. For a free, massless particle wave
packets do not spread, but they do in the other cases.
\end{itemize}
 Effective equations obtained in the way developed here reliably
describe the physical behavior of dynamical wave packets. With these
considerations, effective methods as they have been used in quantum
cosmology are established even in the case of $\phi$-dependent
potentials, as studied specifically for instance in
\cite{BounceSqueezed,Recollapse}. A further application would be to
reconsider the appearance of certain future singularities which have been
shown not to be removed by the tree-level approximation (disregarding
all moments) \cite{FutureSingTree} but where quantum back-reaction is
expected to be strong.

Especially in the presence of potentials, calculations performed
here are much more feasible than the methods involving constructions
of physical Hilbert spaces followed by computations of the
expectation values in explicit physical states. They can be expected
to be of far more general applicability, including full quantum
gravity. For such an extension, several other issues remain open:
describing field theories and handling situations of many classical
degrees of freedom. But there is already a promise that effective
techniques allow one to evade difficult obstacles from physical
inner product issues which so far have impeded progress. Especially
the semiclassical regime of canonical quantum gravity and
potentially observable effects can be brought under much higher
control.

\section*{Acknowledgements}

We thank Don Marolf and Madhavan Varadarajan for discussions.
This work was supported in part by NSF grant PHY0748336.

\appendix

\section{Second order Poisson algebra}
\label{AppI}

\begin{table*}[h]
\centering \caption{Poisson algebra of second order moments. First
terms in the bracket are labeled by rows, second terms are labeled
by columns.}
\begin{tabular}{|c|| c|c|c|c|c|c|c|c|c|c|} \hline & ${\scriptstyle(\Delta t)^2}$
& ${\scriptstyle\Delta(tp_t)}$ & ${\scriptstyle (\Delta p_t)^2}$ &
${\scriptstyle (\Delta q)^2}$ & ${\scriptstyle \Delta(qp)}$ &
${\scriptstyle (\Delta p)^2}$ & ${\scriptstyle \Delta(tq)}$ &
${\scriptstyle \Delta(p_tp)}$ & ${\scriptstyle \Delta(tp)}$ &
${\scriptstyle \Delta(p_tq)}$ \\ \hline\hline ${\scriptstyle (\Delta
t)^2}$ & ${\scriptstyle 0}$ & ${\scriptstyle 2(\Delta t)^2}$ &
${\scriptstyle 4 \Delta(tp_t)}$ & ${\scriptstyle 0}$ &
${\scriptstyle 0}$ & ${\scriptstyle 0}$ & ${\scriptstyle 0}$ &
${\scriptstyle 2 \Delta(tp)}$ & ${\scriptstyle 0}$ & ${\scriptstyle
2 \Delta(tq)}$
\\ \hline ${\scriptstyle \Delta(tp_t)}$ & ${\scriptstyle -2(\Delta
t)^2}$ & ${\scriptstyle 0}$ & ${\scriptstyle 2(\Delta p_t)^2}$ &
${\scriptstyle 0}$ & ${\scriptstyle 0}$ & ${\scriptstyle 0}$ &
${\scriptstyle -\Delta(tq)}$ & ${\scriptstyle \Delta(p_tp)}$ &
${\scriptstyle -\Delta(tp)}$ & ${\scriptstyle \Delta(p_tq)}$ \\
\hline ${\scriptstyle  (\Delta p_t)^2}$ & ${\scriptstyle
-4\Delta(tp_t)}$ & ${\scriptstyle -2(\Delta p_t)^2}$ &
${\scriptstyle 0}$ & ${\scriptstyle 0}$ & ${\scriptstyle  0}$ &
${\scriptstyle  0}$ & ${\scriptstyle  -2\Delta(p_tq)}$ &
${\scriptstyle  0}$ & ${\scriptstyle -2\Delta(p_tp)}$ &
${\scriptstyle 0}$ \\ \hline ${\scriptstyle  (\Delta q)^2}$ &
${\scriptstyle 0}$ & ${\scriptstyle 0}$ & ${\scriptstyle 0}$ &
${\scriptstyle 0}$ & ${\scriptstyle 2(\Delta q)^2}$ & ${\scriptstyle
4\Delta(qp)}$ & ${\scriptstyle 0}$ & ${\scriptstyle 2\Delta(p_tq)}$
& ${\scriptstyle 2 \Delta(tq)}$ & ${\scriptstyle 0}$ \\ \hline
${\scriptstyle \Delta(qp)}$ & ${\scriptstyle 0}$ & ${\scriptstyle
0}$ & ${\scriptstyle 0}$ & ${\scriptstyle -2(\Delta q)^2}$ &
${\scriptstyle 0}$ & ${\scriptstyle 2(\Delta p)^2}$ & ${\scriptstyle
-\Delta(tq)}$ & ${\scriptstyle \Delta(p_tp)}$ & ${\scriptstyle
\Delta(tp)}$ & ${\scriptstyle -\Delta(p_tq)}$ \\ \hline
${\scriptstyle (\Delta p)^2}$ & ${\scriptstyle 0}$ & ${\scriptstyle
0}$ & ${\scriptstyle 0}$ & ${\scriptstyle -4\Delta(qp)}$ &
${\scriptstyle -2(\Delta p)^2}$ & ${\scriptstyle 0}$ &
${\scriptstyle -2\Delta(tp)}$ & ${\scriptstyle 0}$ & ${\scriptstyle
0}$ & ${\scriptstyle -2\Delta(p_tp)}$ \\ \hline ${\scriptstyle
\Delta(tq)}$ & ${\scriptstyle 0}$ & ${\scriptstyle \Delta(tq)}$ &
${\scriptstyle 2\Delta(p_tq)}$ & ${\scriptstyle 0}$ & ${\scriptstyle
\Delta(tq)}$ & ${\scriptstyle 2\Delta(tp)}$ & ${\scriptstyle 0}$ &
${\scriptstyle \Delta(tp_t)}$ & ${\scriptstyle (\Delta t)^2}$ &
${\scriptstyle (\Delta q)^2}$ \\ & & & & & & & & ${\scriptstyle
+\Delta(qp)}$ & &
\\ \hline ${\scriptstyle \Delta(p_tp)}$ & ${\scriptstyle
-2\Delta(tp)}$ & ${\scriptstyle -\Delta(p_tp)}$ & ${\scriptstyle 0}$
& ${\scriptstyle -2\Delta(p_tq)}$ & ${\scriptstyle -\Delta(p_tp)}$ &
${\scriptstyle 0}$ & ${\scriptstyle -\Delta(tp_t)}$ & ${\scriptstyle
0}$ & ${\scriptstyle -(\Delta p)^2}$ & ${\scriptstyle -(\Delta
p_t)^2}$ \\ & & & & & & & ${\scriptstyle -\Delta(qp)}$ & & & \\
\hline ${\scriptstyle \Delta(tp)}$ & ${\scriptstyle 0}$ &
${\scriptstyle \Delta(tp)}$ & ${\scriptstyle 2\Delta(p_tp)}$ &
${\scriptstyle -2\Delta(tq)}$ & ${\scriptstyle -\Delta(tp)}$ &
${\scriptstyle 0}$ & ${\scriptstyle -(\Delta t)^2}$ & ${\scriptstyle
(\Delta p)^2}$ & ${\scriptstyle 0}$ & ${\scriptstyle \Delta(qp)}$ \\
& & & & & & & & & & ${\scriptstyle -\Delta(tp_t)}$ \\ \hline
${\scriptstyle \Delta(p_tq)}$ & ${\scriptstyle -2\Delta(tq)}$ &
${\scriptstyle -\Delta(p_tq)}$ & ${\scriptstyle 0}$ & ${\scriptstyle
0}$ & ${\scriptstyle \Delta(p_tq)}$ & ${\scriptstyle 2\Delta(p_tp)}$
& ${\scriptstyle -(\Delta q)^2}$ & ${\scriptstyle (\Delta p_t)^2}$ &
${\scriptstyle \Delta(tp_t)}$ & ${\scriptstyle 0}$ \\ & & & & & & &
& & ${\scriptstyle -\Delta(qp)}$ & \\ \hline
\end{tabular}
\label{tab:pb_moments}
\end{table*}

The expectation values obey the classical Poisson algebra for two
canonical pairs, where the non trivial brackets are
\[
\{ t, p_t \} = 1 \quad {\rm and} \quad \{ q, p\} = 1\,.
\]
The Poisson brackets between the expectation values and the moments
are zero. In Table~\ref{tab:pb_moments} we provide the Poisson
brackets for the second order moments of quantum variables
associated with two canonical pairs $\mathbf{t},\mathbf{p}_t;
\mathbf{q},\mathbf{p}$.

\section{Coherent state and effective evolution}
\label{AppII}

Here we consider the constrained system of Section~\ref{sec:space_pot}
in a specific semiclassical state and compare the evolution of
``classical'' quantities given by the effective quantum theory to
properties of the state. This appendix is included to address the
riliability of an effective solution to a constrained quantum system
through comparison within a specific example.

Recall the constraint:
\[
\mathbf{C} = \mathbf{p}_t^2 - \mathbf{p}^2 - \mathbf{q}^2 - m^2
\mathds{1}\,.
\]
Formally, a positive frequency solution of the constraint described
in Section~\ref{sec:space_pot} reduces the system to one canonical
degree of freedom that evolves subject to the Hamiltonian
$\mathbf{H} = \left( \mathbf{p}^2 + \mathbf{q}^2 +
m^2\mathds{1}\right) ^{\frac{1}{2}}$. We begin by providing the
classical trajectory: a canonical pair $q$, $p$\ subject to the
Hamiltonian function $H = \left( p^2 + q^2 + m^2\right)
^{\frac{1}{2}}$\ evolves according to the equations of motion:
\begin{align*}
\frac{\rm d \ }{{\rm d} t} q &= \{q, H\} =  p \left( p^2 + q^2 +
m^2\right)^{-\frac{1}{2}} = pH^{-1} \\ \frac{\rm d \
}{{\rm d} t} p &= \{p, H\} = -q \left( p^2 + q^2 +
m^2\right)^{-\frac{1}{2}} = qH^{-1}\,.
\end{align*}
Using the fact that $H$\ itself is a constant of motion we
differentiate the first equation above with respect to time to
obtain the second order equation
\[
\frac{\rm d^2 \ }{{\rm d} t^2} q = H^{-1} \frac{\rm d \ }{{\rm d} t}
p = -H^{-2} q\,.
\]
This equation has the general solution
\[
q(t) = A \sin \left( \frac{t}{H} \right) + B \cos \left( \frac{t}{H}
\right)\,,
\]
where $A$\ and $B$\ are constants.
It follows that
\[
p(t) = H \frac{\rm d \ }{{\rm d} t} q = A \cos \left( \frac{t}{H}
\right) - B \sin \left( \frac{t}{H} \right)
\]
and thus $H = \sqrt{A^2 + B^2 + m^2}$.
The classical phase-space trajectory is therefore a circle of radius
$\sqrt{A^2+B^2}$\ traversed with the angular frequency $\left( A^2 +
B^2 + m^2 \right)^{-\frac{1}{2}}$.

Effective equations of motion are set up in much the same way as
their classical counterparts. Phase-space degrees of freedom are
$q$, $p$, $(\Delta q)^2$, $(\Delta p)^2$, $\Delta(qp)$. The time
evolution is generated by the function $E$\ of
equation~(\ref{eq:E_space_pot}) through the quantum Poisson bracket
\begin{align*}
\frac{\rm d \ }{{\rm d} t} q &= \frac{p}{\sqrt{p^2 + q^2 + m^2}} +
\frac{p (\Delta q)^2 \left( 2q^2 - p^2 - m^2 \right) + q \Delta(qp)
\left( 4p^2 - 2q^2 - 2m^2 \right) - 3p (\Delta p)^2 \left( q^2 + m^2
\right)}{2 \left( p^2 + q^2 + m^2 \right)^{\frac{5}{2}}} \\
\frac{\rm d \ }{{\rm d} t} p &= \frac{-q}{\sqrt{p^2 + q^2 + m^2}} +
\frac{3q (\Delta q)^2 \left( p^2 + m^2 \right) - p \Delta(qp) \left(
4q^2 - 2p^2 - 2m^2 \right) - q (\Delta p)^2 \left( 2p^2 - q^2 - m^2
\right)}{2 \left( p^2 + q^2 + m^2 \right)^{\frac{5}{2}}} \\
\frac{\rm d \ }{{\rm d} t} (\Delta q)^2 &= \frac{2\Delta(qp) \left(
q^2 + m^2 \right) - 2(\Delta q)^2qp}{\left( p^2 + q^2 + m^2
\right)^{\frac{3}{2}}} \quad \quad , \quad \quad \frac{\rm d \
}{{\rm d} t} (\Delta p)^2 = \frac{2 (\Delta p)^2qp - 2\Delta(qp)
\left( q^2 + m^2 \right)}{\left( p^2 + q^2 + m^2
\right)^{\frac{3}{2}}} \\
\frac{\rm d \ }{{\rm d} t} \Delta (qp) &= \frac{ (\Delta p)^2 \left(
q^2 + m^2 \right) - (\Delta q)^2 \left( p^2 + m^2 \right)}{\left(
p^2 + q^2 + m^2 \right)^{\frac{3}{2}}}\,.
\end{align*}
The first term in each of the first two equations is identical to
the classical equations of motion, the extra terms constitute the
leading order quantum corrections. The system of equations is
straightforward to evolve numerically for a sufficiently short time
starting from a state that is initially semiclassical.

The quantum evolution of positive frequency solutions is governed by
the Schr\"odinger equation
\[
-\frac{\hbar}{i} \frac{\rm d}{{\rm d} t} \Psi(x, t) = \left(
\mathbf{p}^2 + \mathbf{q}^2 + m^2\mathds{1} \right)^{\frac{1}{2}}
\Psi(x, t)\,.
\]
The square root operator has the same eigenstates $\{
\varphi_n(x)\}_{n=0, \ldots \infty;}$\ as the one dimensional
quantum harmonic oscillator, with eigenvalues $\lambda_n =
\sqrt{2(n+\frac{1}{2})\hbar + m^2}$. A wavefunction can be evolved
by decomposing it into the eigenstates of the Hamiltonian. If the
state at $t=0$\ is given by $\Psi(x, 0) = \sum_{n=0}^{\infty} c_n
\varphi_n(x)$, where $c_n$\ are constant complex numbers, then at
any other time, the wavefunction is
\begin{equation}\label{Psixt}
\Psi(x, t) = \sum_{n=0}^{\infty} c_n \exp \left( -i \frac{\lambda_n
t}{\hbar} \right) \varphi_n(x)\,.
\end{equation}
To compute the expectation values we write $\mathbf{q} =
\sqrt{\hbar/2} \left( \hat{a}^{\ast} + \hat{a} \right)$,
$\mathbf{p} = i\sqrt{\hbar/2} \left( \hat{a}^{\ast} -
\hat{a} \right)$, where $\hat{a}^{\ast}$\ and $\hat{a}$\ are the
usual creation and annihilation operators of the quantum harmonic
oscillator. One finds
\begin{align*}
\langle \Psi, \mathbf{q} \Psi \rangle (t) =& \sum_{n=0}^{\infty}
\sqrt{2\hbar(n+1)} {\rm Re} \left[ \bar{c}_n
c_{n+1} \exp \left( -it \frac{\lambda_{n+1} - \lambda_{n}}{\hbar}
\right) \right] \\ \langle \Psi, \mathbf{p} \Psi \rangle (t) =&
\sum_{n=0}^{\infty} \sqrt{2\hbar(n+1)} {\rm Im}
\left[ \bar{c}_n c_{n+1} \exp \left( -it \frac{\lambda_{n+1} -
\lambda_{n}}{\hbar} \right) \right]
\end{align*}
In a similar manner one can obtain expressions for the moments of
$\mathbf{q}$\ and $\mathbf{p}$.

\begin{figure}[htbp!]
\includegraphics{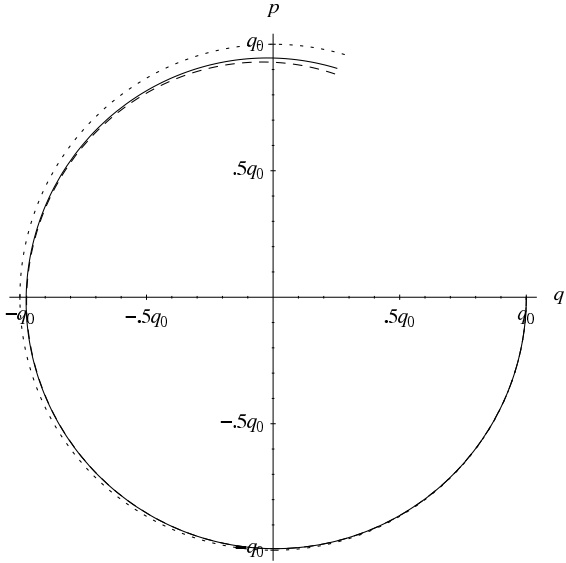}
\caption{\label{fig:1} Classical (dotted), coherent state (solid) and
effective (dashed) phase space trajectories, evolved for $0 \leq t
\leq 5q_0$.}
\end{figure}

In order to complete the comparison we select a semiclassical state
with a known decomposition into the eigenstates $\{ \varphi_n (x)
\}$. A simple choice is to set the initial wavefunction to be a
coherent state of the harmonic oscillator
\begin{equation}\label{cn}
c_n = \exp \left( -\frac{|\alpha|^2}{2} \right)
\frac{\alpha^n}{\sqrt{n!}}, \quad \alpha \in \mathds{C}
\end{equation}
which we can consider as a kinematical coherent state for our
system. For the non-relativistic harmonic oscillator, as time goes
on $\alpha$ changes, but the shape of the state is preserved.
Clearly, this is not the case for our relativistic evolution:
Combining (\ref{cn}) with (\ref{Psixt}), we have an evolving state
which, expanded in harmonic oscillator stationary states, has
coefficients
\[
 c_n e^{-i\lambda_n t/\hbar}=\frac{1}{\sqrt{n!}}e^{-|\alpha|^2/2}\alpha^n
e^{-i\sqrt{2n+1+m^2/\hbar^2}\,t}\,.
\] Due to the square root in this
relativistic model, these coefficients are not of the coherent state
form (\ref{cn}) unless $t\not=0$. The physical states we obtain are
not dynamical coherent states; quantum back-reaction ensues which in
the effective treatment is captured by the coupling terms between
moments and expectation values in (\ref{eq:E_space_pot}).

\begin{figure}[htbp!]
\includegraphics{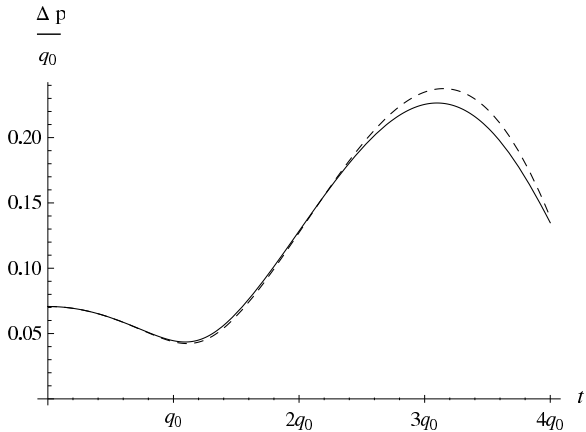}
\caption{\label{fig:2} Coherent state (solid) and effective (dashed)
evolution of the second order moment $\Delta p= \sqrt{(\Delta
p)^2}$\ in units of $q_0$.}
\end{figure}

For a specific numerical example, we set $\alpha =
\frac{q_0}{\sqrt{2\hbar}}$\ so that at $t=0$\ the state is a Gaussian
peaked about $q = q_0$\ and $p = 0$, with zero covariance and minimal
spread
\[
(\Delta q)^2 = (\Delta p)^2 = \frac{\hbar}{2}, \quad \Delta(qp) =
0\,.
\]
We take these as initial values for the numerical evolution of the
effective equations for the system. We assume the two physical scales
to be separated by a single order of magnitude by setting
$\frac{q_0}{\sqrt{\hbar}} = 10$. For simplicity we set $m=0$. Depicted
in FIG.~\ref{fig:1} are the classical, coherent and effective quantum
phase-space trajectories starting from the same initial state. One can
see that the effective equations describe the correct semiclassical
trajectory for much of the evolution displayed. An internal measure of
consistency is the size of second order moments.  From
FIG.~\ref{fig:2}, we see that the semiclassical approximation clearly
breaks down after approximately $t=2q_0$, as $(\Delta p)^2$\ becomes
too large. The same figure demonstrates that up until that time the
evolution of the moments themselves is very well approximated by the
effective equations. The other two moments display similar behavior.

\end{document}